\documentstyle[psfig,conf_iap,10pt]{article}

\def\kms{\ifmmode\,{\rm km}\,{\rm s}^{-1}\else km$\,$s$^{-1}$\fi}
\def\ifm#1{\relax\ifmmode#1\else$\mathsurround=0pt #1$\fi}
\def\hmpc{\,h\ifm{^{-1}}{\rm Mpc}}
\def \etal{{\sl et al.\ }}

\def \fm{\hbox{$.\!\!^{\rm m}$}}

\def \gtsima{$\, \buildrel > \over \sim \,$}
\def \ltsima{\, \buildrel < \over \sim \,}
\def \simgt{\lower.5ex\hbox{\gtsima}}
\def \simlt{\lower.5ex\hbox{\ltsima}}

\begin{document}

\heading{SHELLFLOW: The First Homogeneous All-Sky TF \\ Survey at 6000 \kms} 

\par\medskip\noindent

\author{
St\'ephane Courteau$^{1}$, Jeff Willick$^{2}$, Michael Strauss$^{3}$, 
David Schlegel$^{3}$, Marc Postman$^{4}$
}

\address{Herzberg Institute of Astrophysics, Victoria, BC}

\address{Stanford University, Department of Physics, Stanford, CA}

\address{Princeton University, Department of Astrophysical Sciences, 
 Princeton, NJ}

\address{Space Telescope Science Institute, Baltimore, MD}

\begin{abstract}
We present a new optical Tully-Fisher (TF) investigation for a complete, 
full-sky sample of 297 Sb$-$Sc spirals with redshifts between 4500 and 
7000\kms.  The survey was specifically designed to provide {\it uniform,
well-calibrated} data over both hemispheres.  All previous TF surveys
within the Supergalactic shell ($cz \ltsima 6000$ \kms) have relied on 
matching separate data sets in the Northern and Southern hemispheres 
and thus cannot attain full-sky homogeneity.  Analyses of 
the cosmological dipole and peculiar velocities based on these studies
have produced contradictory claims for the amplitude of the bulk flow 
and whether it is generated by internal or external mass fluctuations. 
With Shellflow, and further zero-point calibration of existing TF data 
sets, we expect a high-accuracy detection of the bulk flow amplitude 
and an unambiguous characterization of the tidal field at 6000\kms.


\end{abstract}

\section{Introduction}
Peculiar velocity surveys covering a fair fraction of the sky are now reaching
to 6000 \kms\ and beyond (\cite{C92}, \cite{MAT92}, \cite{MAT94}, \cite{da96},
\cite{Hu98}, \cite{W98}) and are being interpreted as evidence for substantial 
flows on these scales (\cite{C93}, \cite{MAT94}, \cite{LP94}, \cite{da96}, 
\cite{RPK}, \cite{SW95}). However, the amplitude, direction, and scale of 
these flows remain very much in contention, with resulting uncertainties in 
the theoretical interpretation and implications of these measurements 
(\cite{Po95}, \cite{SW95}).  

Indeed, recently published conflicting results suggest that the motion of 
the LG is either due, or is not due, to material within 6000 \kms, and that 
{\sl IRAS\/} galaxies either trace, or do not trace, the dark matter which 
gives rise to the observed peculiar velocities.

The most recent POTENT reconstruction of the MarkIII velocities (\cite{DK99})
shows that the bulk velocity can be decomposed into two components arising 
from the mass fluctuation field within the sphere of radius $60\hmpc$ about
the LG and a component dominated by the mass distribution outside that 
volume.  For convenience, we refer to this boundary at $60\hmpc$ as the 
``Supergalactic shell'' since it includes the main local attractors in 
the Supergalactic plane, the Great Attractor and Perseus-Pisces.  This new
analysis shows dominant infall patterns by the GA and PP but very little 
bulk flow within the Supergalactic shell.  The tidal component inside this 
volume is dominated by a flow of amplitude $370 \pm 125 \kms$ in the 
Supergalactic direction $(L,B)=(165^\circ,\, -10^\circ)$, which is likely 
generated by the external mass distribution on very large scales (see also 
\cite{CB81}, \cite{C93}).  This interpretation is also supported by an 
increasingly large number of TF/FP investigations (based on the distribution 
and motion of Abell clusters) which report the detection of streaming motions 
of amplitudes greater than 700 \kms beyond $\sim 60\hmpc$ and away from 
the CMB dipole (\cite{Sca89}, \cite{LP94}, \cite{W98}, \cite{Hu98}).  
Other investigations using nearly homogeneous samples of galaxies within and 
outside the Supergalactic shell find motion consistent with the amplitude 
and direction of the CMB dipole {\cite{Gio98}}.  This suggests that 
the reflex motion of the Local Group could be explained by material 
contained within the Supergalactic shell. 

This confusion stems, in large part, in our inability to perfectly match 
the many heterogeneous samples for flow studies into one self-consistent 
homogeneous catalogue.  Much of the problem lies in the fact that, with 
the exception of a few surveys beyond $\sim 100\hmpc$ (\cite{LP94}, 
\cite{W98}, \cite{Hu98}), none of the surveys within the Supergalactic 
sphere sample the {\it entire} sky uniformly.  

\section{The MarkIII Catalog of Galaxy Peculiar Velocities}

In an attempt to overcome this problem, two of us (JW \& SC $+$ collaborators)
have recently combined the major distance-redshift surveys from both 
hemispheres (published before 1994) into a catalog of 3100 galaxies
(\cite{W98}), but showed that full homogenization at the $2$\% level, the 
minimum required for a $\geq 3 \sigma$ bulk flow detection at 6000 \kms, 
cannot be achieved.  Due to subjective reduction techniques and varying
selection criteria, fundamental uncertainties remain when trying to match 
greatly disparate TF datasets (\cite{W97}).
Furthermore, a revised calibration of the MarkIII TF zero-points 
based on maximal 
agreement with the peculiar velocities predicted by the IRAS 1.2Jy redshift 
survey suggests a possible source of systematic error for the
data sets which cover the PP cone (\cite{WS98}).  This uncertainty has
not seriously affected mass density reconstructions within the Supergalactic
shell (\cite{DK99}) but it could lead to spurious estimates of the bulk 
flows on larger scales.  A newer calibration of the Courteau/Faber catalogue 
of Northern spirals, not included in MarkIII, has been published 
(\cite{C96}, \cite{C97}) but a revision of the MarkIII catalogue is in 
progress (\cite{W99}).

\section{A New All-Sky Survey: SHELLFLOW}

The need to tie all existing data bases for cosmic flow studies in an
unambiguous fashion is clear.  To that effect, we initiated a 
new survey in 1996 using NOAO facilities to measure TF distances for a 
complete, full-sky sample of Sb$-$Sc galaxies in the Supergalactic shell
for which we will obtain {\it precise\/} and {\it uniform\/} photometric 
and spectroscopic data.  This will be the first well-defined full-sky 
survey to sample this scale, free of uncertainties from matching 
heterogeneous data sets.  
The SFI survey of Giovanelli \etal \cite{Gio98} resembles ours in its 
scope and sky coverage, but it relies on a separate dataset (\cite{MAT92}) 
for coverage of the Southern Sky and thus cannot attain full-sky homogeneity.
Our survey, on the other hand, is designed from the outset to be homogeneous 
to the minimum level required for unambiguous bulk flow detection at the 
Supergalactic shell.  Because of the overlap with existing surveys at 
comparable depth (MarkIII + SFI), this new compilation will be of fundamental 
importance in tying the majority of existing data sets together in a uniform 
way, which will greatly increase their usefulness for global analyses of 
mass fluctuations in the universe.  

Our sample is selected from the Optical Redshift Survey (\cite{SS95}),
consisting of galaxies over the whole sky with m$_{\rm{B}} \geq 14.5$
and $|b| \geq 20^\circ$ from the UGC, ESO, and ESGC (\cite{Cor}).
It includes all non-interacting Sb and Sc galaxies with redshifts between 
4500 and 7000 \kms\ from the Local Group and inclinations between 
$45^\circ$ and $78^\circ$, in regions where Burstein-Heiles extinction 
is less than 0\fm3. This yields an all-sky catalog of 297 galaxies.
Following the approach of \cite{LP94}, we use the sample itself to 
calibrate the distance indicator relation; this mitigates the need to 
tie the sample to external TF calibrators such as clusters (although
it precludes measurement of a monopole term in the velocity field).
Given a TF fractional distance error of 20\%, the statistical uncertainty 
on a bulk flow from $N=297$ galaxies at common distance $D = 6000 \kms$ 
is $\Delta D/\sqrt{N} = 70 \kms$. As the measured (and much contested) 
bulk motions on these scales are of the order of 300 \kms, a detection 
of high statistical significance is well within reach.

\section{Results and Analysis}
Data taking and reduction techniques follow the basic guidelines of previous
optical TF surveys (\cite{C96}, \cite{Sch96}, \cite{C97}, \cite{W98}).  
Our survey is now complete, which is essential to achieve our statistical 
requirements and ensure a rigorous analysis.  The spectroscopy relies on 
measurement of H$\alpha$ rotation velocities at 2.2 disk scale lengths 
for the tightest TF calibration and best match to analogous 21cm line 
widths (\cite{C97}, \cite{W98}).  The photometry is based on the 
Kron-Cousins $V$ and $I$ systems which will allow direct matching 
with two largest TF field samples to date (\cite{MAT92},\cite{Gio98}).
One of the key features of this study is not only its all-sky sample 
selection but the independent duplication of all data reductions 
(by at least 2, if not 3, of us).  These reductions and a first flow 
analysis based on the Shellflow sample alone should be published soon 
(\cite{C99}).  We also plan a more 
extensive analysis using the recalibrated MarkIII combined with other new 
catalogs not included in the original MarkIII. 








\vfill
\end{document}
\end